\documentclass[english,12pt, oribibl]{article}
\usepackage[T2A]{fontenc}
\usepackage{amsbsy}
\usepackage{amsmath}
\usepackage[english]{babel}
\usepackage{xcolor}
\usepackage{graphicx}
\usepackage{hyperref}
\usepackage{geometry} 
\date{}
\title{THERMODIFFUSION UNIPOLAR ELECTRIC GENERATOR}

\newcommand{\footremember}[2]{%
	\footnote{#2}
	\newcounter{#1}
	\setcounter{#1}{\value{footnote}}%
}
\newcommand{\footrecall}[1]{%
	\footnotemark[\value{#1}]%
} 
\def\keywords{\vspace{.5em}
	{\textit{Keywords}:\,\relax%
}}

\author{%
	G. S. Bisnovatyi-Kogan\footremember{one}{Space Research Institute Russian Academy of Sciences}, \footremember{two}{NRNU MEPhI}%
	\and M. V. Glushikhina\footrecall{one}
}
\begin{document}
	\maketitle
	
	\begin{abstract}
	A model of a conducting cylinder with a radial temperature gradient which creates an electric field that
	increases with time in the surrounding vacuum is examined. The conditions under which this model
	functions are pointed out. An electric field is also generated when a magnetic field exists along the axis
	of the cylinder. This article discusses the interactions of the thermal flux, magnetic field, and charge
	distribution. Four models are considered with different conditions for the supply of electrons from a
	central source and the possibility either of capturing electrons inside the cylinder or their freely leaving
	it through the outer boundary.	
	\end{abstract}
	


		\section{Introduction}

		Generation of an electric field by a rotating unipolar inductor is a well-known classical electrodynamic effect \cite{Tamm-1976}.
		In space objects, it has been observed by astronomers since the middle of the last century.  The idea of accelerating charged particles by rotating magnetized stars as unipolar inductors was first proposed by Yu. Terletsky in 1945. \cite{terl45}.
		Later, several authors considered the possibility of accelerating charged particles by the unipolar induction field of the Sun and stars to explain the origin of cosmic rays\,\cite{Terletsky-1946} $^-$\, \cite{ginzburg1953}.

		After the discovery of radio pulsars   
		P. Goldreich and W. Julian \cite{Goldreich-Julian-1969} applied this mechanism to describe processes in the pulsar magnetosphere. Currently, this mechanism is considered one of the most important for the acceleration of electrons and protons, the loss of rotational energy of magnetized neutron stars, and the formation of pulsar wind nebulae (PWN). 
		
		Unipolar induction was considered as
		an effective mechanism for extracting
		rotational energy from a rotating Kerr black hole surrounded by highly magnetized plasma. Various names were proposed for this effect:
		black hole dynamo \cite{Znajek-1978},
		unipolar induction battery \cite{Phinney-1982},
		surface battery \cite{Thorne-1986}.
		The problems arising from the application of the unipolar induction mechanism in the magnetized plasma surrounding a Kerr black hole were discussed in the article of Okamoto  \cite{Okamoto-2015}.
		
		The effect of the unipolar mechanism on Jupiter's moon Io was considered by Goldreich and Lyndon-Bell in 1969 \cite{Goldreich-Lynden-Bell}. They suggested that the flow of electrons accelerated by this mechanism from Io causes the observed bursts of decameter emission from Jupiter when interacting with its ionosphere.  	
		
		The magnetic field of a linear current surrounding  a perfectly conducting cylinder with a gap was considered by M.A. Leontovich \cite{Leontovich-1952}. The properties of the plasma cylinder for various conditions were studied by I.E. Tamm \cite{Tamm-1951}.  These works were carried out in connection with the start of work on the creation of a thermonuclear reactor, which is still under construction.

		In this article, we propose  mechanism for unipolar induction in which a conductive cylinder with a radial temperature gradient plays the role of a rotating magnetized cylinder \cite{Tamm-1976}.
		Considering a simplified cylindrical model in a homogeneous magnetic field and without it, we study the interaction of thermal flux, magnetic field and charge distribution in this model.   The greatest attention is paid to the variant with a growing electric field in a vacuum.

		The presence of a radial heat flux, which leads to the creation of a radial electric current due to the effect of thermal diffusion, as well as the presence of an azimuthal magnetic field created by a possible electric current along the axis, does not violate cylindrical symmetry. In the laboratory, the appearance of an electric current along the axis of the cylinder can be avoided by using two identical flows of electrons moving oppositely. In the following presentation, the absence of a longitudinal current from external sources $j_z=0$ is everywhere accepted, and the inertia of electrons in macroscopic phenomena is also neglected.

		\section{Magnetic fields and electric currents in a conductive cylinder}
		\label{2}
		
		Consider a cylinder, presumably with a non-degenerate, non-relativistic plasma, with zero velocity of matter, with a temperature gradient directed along the radius, with a possible uniform magnetic field $B$ along the $Z$ axis. It is assumed that the only source of heat and electrons is located near the axis of the cylinder and is represented by a uniformly heated cylinder with a radius of $R_1 << R_0$, where $R_0$ is the outer radius of the cylinder, $R_1$ is the radius of the inner cylinder, the inner "string".
		
		The transfer coefficients determining heat flux and diffusion
		in plasma have a tensor structure in a magnetic field. This means that the direction of the thermal and diffusion fluxes does not coincide with the direction of the corresponding electric field vectors $\bf E$ and the temperature gradient $\nabla\bf T$ responsible for the formation of these fluxes. 
		Part of the electric current vector $\textbf{j}$ is connected to the electric field vector ${\bf E}$, which makes up the main part of the diffusion vector ${\textbf{d}}$, using the electrical conductivity tensor $\overleftrightarrow{\sigma_E}$. The other part of ${\textbf{j}}$ is related to the temperature gradient vector $\nabla\bf T$ by the tensor $\overleftrightarrow{\sigma_T}$.

		In \cite{BK-Glush-2018,BK-Glush-2018a,Glu-2020}, the components of four tensors of kinetic coefficients, namely, thermal conductivity, diffusion, thermodiffusion and diffusion thermal effect - were calculated for various conditions, including highly degenerate plasma.

		The equations for the heat flux $q_i$ and the diffusion rate $v_i$ are written as follows:
		
		\begin{eqnarray}\label{q_ii}
			q_i=q_{i}^{(T)}+q_{i}^{(D)}=
			-\left(\lambda^{(1)}\delta_{ij}-\lambda^{(2)}\varepsilon_{ijk}B_k+\lambda^{(3)}B_i B_j\right)\frac{\partial T}{\partial x_j} \nonumber\\ -\frac{e n_e}{kT }\left(\nu^{(1)}\delta_{ij}-\nu^{(2)}\varepsilon_{ijk}B_k+\nu^{(3)}B_i B_j\right)E_j,
		\end{eqnarray}
		
		\begin{eqnarray}\label{v_ii}
			\langle v_{i} \rangle=\langle v_{i}^{(D)} \rangle + \langle
			v_{i}^{(T)} \rangle
			=  -\frac{e n_e}{kT} \left(\eta^{(1)}\delta_{ij}
			-\eta^{(2)}\varepsilon_{ijk}B_k+\eta^{(3)}B_i B_j\right)E_j \\
			-\left(\mu^{(1)}\delta_{ij}-\mu^{(2)}\varepsilon_{ijk}B_k
			+\mu^{(3)}B_i B_j\right)\frac{\partial T}{\partial x_j},\nonumber
		\end{eqnarray}

		Indices (T) and (D) correspond to the heat flux and the electron diffusion rate determined by the temperature gradient $\partial T/\partial x_j$ and the electric field $E_i$ (as part of the diffusion vector), respectively.

		Kinetic coefficients
		$\lambda^{(i)}$, $\mu^{(i)}$, $\eta^{(i)}$ and $\nu^{(i)}$ determine thermal and diffusion fluxes
		in the following directions. The upper indices $^{(1)}$
		define the above mentioned flows along the temperature gradient
		$\partial T/\partial x_i$, or the diffusion vector $d_i$.
		The upper indices $^{(3)}$ are related to the direction along the magnetic field; and the upper indices $^{(2)}$ define flows perpendicular to the plane defined by the magnetic field vector $B_i$ and any of the vectors $\partial T/\partial x_i$ or $d_i$. These flows are called Hall $q_{Hall}$ and $j_{Hall}$.

		Plasma in the crusts of neutron stars is collisional due to its high density.  The electron gas there is in the crystal lattice of heavy nuclei that the electrons collide with. 
		
		One of the first calculations of kinetic coefficients in a magnetized plasma by solving the Boltzmann equation using the Chapman-Enskog  method \cite{Chapmen-1952} was carried out by	Braginsky \cite{Brag-1958}, \cite{Brag-1963}.
		The kinetic coefficients for a fully ionized magnetized plasma were calculated by direct numerical calculation of the Fokker-Planck equation in \cite{Epperlein-Haines-1986}. In the works \cite {BK-Glush-2018}, \cite{Glu-2020}, analytical expressions for four tensor kinetic coefficients in a magnetized plasma were obtained by solving the Boltzmann equation in a three-polynomial approximation, as a refinement of the two-polynomial results \cite{Brag-1958},\cite{Brag-1963}.

		We use kinetic coefficients obtained by the Lorentz  method \cite{Chapmen-1952}. In this approximation, the kinetic coefficients in the absence of a magnetic field are calculated from the exact solution of the linearized Boltzmann equation \cite{Chapmen-1952}, \cite{schatz-1958}, \cite{BK-2001}.
		The errors of the results in the Lorentz approximation do not exceed 50\%, which is sufficient for a qualitative description of the processes in the problem under consideration.

	With cylindrical symmetry $\frac{\partial}{\partial z}=\frac{\partial}{\partial\phi}=0$, the only nonzero values remain: $q_r,\,\,q_\phi,\,\,j_r,\,\,j_\phi,\,\,B_z,\,\,E_r$. For non-degenerate electrons from (\ref{q_ii}),(\ref{v_ii}) using definitions of electric current density:
	\begin{equation}
		\label{ji}
		j_i = -n_e e \langle v_i \rangle,
	\end{equation}
	in the presence of a magnetic field along the axis of the cylinder, we have the following relations:
	\begin{eqnarray}
		\label{qrp}
		q_r=-\lambda^{(1)}\frac{dT}{dr}-\frac{e n_e}{kT}\nu^{(1)}E_r,\\ \nonumber
		{q_\phi=-B_z  \left(\lambda^{(2)}\frac{dT}{dr}+\frac{e n_e}{kT}\nu^{(2)}E_r\right),}
	\end{eqnarray}
	\begin{eqnarray}
		\label{vr}
		v_r=-\mu^{(1)}\frac{dT}{dr}-\frac{e n_e}{kT}\eta^{(1)}E_r, \\ \nonumber
		v_\phi=-B_z  \left(\mu^{(2)}\frac{dT}{dr}+\frac{e n_e}{kT}\eta^{(2)}E_r\right),
	\end{eqnarray}
	
	\begin{eqnarray}
		\label{jr}
		j_r=e n_e\left(\mu^{(1)}\frac{dT}{dr}+\frac{e n_e}{kT}\eta^{(1)}E_r\right), \\ \nonumber
		j_\phi=B_z  en_e\left(\mu^{(2)}\frac{dT}{dr}+\frac{e n_e}{kT}\eta^{(2)}E_r\right).
	\end{eqnarray}
	
	The coefficients of thermal conductivity and thermal diffusion, calculated in the Lorentz approximation for the case of a zero magnetic field, are written as:
	\begin{equation}
		\label{tlor}
		\tilde\lambda_T =   \frac{320}{3\pi}\frac{k^2 Tn_e}{m_e}\tau_e,
	\end{equation}
	
	\begin{equation}
		\label{mulor}
		\mu^{(1)} =  \frac{16k}{m_e \pi}\tau_e \equiv \frac{\sigma_T}{en_e}.
	\end{equation}
	Using the expression for the electric current density, you can write the part of it associated with the temperature gradient using the thermal diffusion coefficient in the form:
	
	\begin{equation}
		j_r^T = -n_e e \langle v_r^T \rangle  ={- n_e e (-
			\mu^{(1)} \frac{dT}{dr})} =\sigma_T \frac{dT}{dr}
	\end{equation}
	
	The following plasma parameters are used here and further: Larmor electron frequency $\omega_B=\frac{eB}{m_e c}$, time between $eN$ collisions $\tau_e =\dfrac{3}{4}\sqrt{\dfrac{m_e}{2\pi}}\dfrac{(kT)^{3/2}}{Z^2{e}^4n_N\Lambda}$,
	the coefficient of thermal conductivity $\sigma_T$, which for the approximation of a non-degenerate Lorentz gas is defined in \cite{BK-2001}; $n_e,\,\,n_N$ is the concentration of electrons and nuclei with atomic number $Z$, $\Lambda$ is the Coulomb logarithm.
	
	The coefficients of electrical conductivity $\sigma_E$ and diffusion thermal effect $\lambda_E$ are written as
	
	\begin{equation}
		\sigma_E= \frac{e^2 n^2_e}{kT} \eta^{(1)}=\frac{32}{3\pi}
		\frac{e^2 n_e}{m_e}\tau_{e} ,\\ \quad
		\lambda_E= \frac{n_e}{kT} \nu^{(1)}=\frac{128}{3\pi}
		\frac{e kT n_e}{m_{e}}\tau_{e}.
	\end{equation}
	Thus, the heat flux $q_i$ and the electric current $j_i$ in
	an unmagnetized plasma are written as:
	\begin{equation}
		\label{qj1}	
		q_r= -\tilde\lambda_T \frac{dT}{dr} - \lambda_E E_r,\\ \quad
		j_r=\sigma_T \frac{dT}{dr}+ \sigma_E E_r.
	\end{equation}
	
	In the presence of a magnetic field along the axis of the cylinder, we need
	to use one of Maxwell's equations to calculate the additional magnetic field $B_z$ created by the azimuthal electric current arising from the effect
	Hall \cite{bkglaph23}. In the absence of a longitudinal current, we will write down the components we need in the following form:
	\begin{eqnarray}\label{maxwell-general}
		rot \textbf{B} = \frac{4 \pi}{c} \textbf{j}, \qquad \qquad \qquad \qquad \qquad \qquad \qquad \\ \nonumber 
		\quad \frac{dB_r}{d z} - \frac{dB_{z}}{dr} = \frac{4 \pi}{c} j_{\phi}, \quad 
		\frac{1}{r}\frac{d }{dr}(r B_{\phi}) - \frac{1}{r}\frac{dB_{r}}{d\phi} = \frac{4 \pi}{c} j_z. 
	\end{eqnarray}
	At the same time, it is necessary to distinguish between external electromotive forces associated with a temperature gradient and electric forces acting inside a conductive medium, see the discussion of this issue in the book by I. E. Tamm\cite{Tamm-1976}.   	
	The nonzero components of the electric current density vector $j_i$ in the cylinder are determined as follows \cite{Chapmen-1952}, 
	
	\begin{eqnarray}\label{eq2}
		j_r=\frac{\sigma_T ({\bf \nabla}T)_r+\sigma_E E_r}{1+\omega_B^2\tau_e^2}, \qquad\qquad\\
		j_\varphi=\frac{(\sigma_T\,({\bf \nabla}T)_r +
			\sigma_E E_r) \omega_B\tau_e }{1+\omega_B^2\tau_e^2},\quad j_z=0.\nonumber
	\end{eqnarray}
	Taking into account the symmetry of the cylinder, the stationarity of our model and the equations (\ref{maxwell-general}) and (\ref{eq2}), we write the Maxwell equation\eqref{maxwell-general} as:
	
	\begin{equation}
		B_r=B_\varphi=0,\quad \frac{c}{4\pi}\frac{d B_z}{d r}=\frac{(\sigma_T\,({\bf \nabla}T)_r+ \sigma_E E_r)\omega_B\tau_e }{1+\omega_B^2\tau_e^2}.
		\label{maxwell}
	\end{equation}
	The "Hall" component of the magnetic field $B_z$ defined in this way is an addition (presumably small) to the initial longitudinal magnetic field, which determines the Larmor frequency $\omega_B$.  	
	
	The main purpose of this work is to build a model
	of an electric field generator in a plasma or metal cylinder with a radial temperature gradient. The Lorentz approximation for kinetic coefficients is used. Deviations from more precise formulas in plasma are about several tens of percent, but using them, we get a simple analytical solution to the equation. 
	In the crust of a neutron star, where matter is in the Coulomb crystal state, the results of the Lorentz approximation are almost accurate.

	The relationship of the vectors ${\bf j}$ and ${\bf B}$, determined by the Maxwell equation, is considered for two different conditions on the inner and outer boundaries of the cylinder.

	\indent I. The electrons cross the outer boundary without any supply of electrons from the center. The stationary state
	is established with a positive charge of the cylinder, zero electric current and a non-zero radial electric field $E_r(r)$, both inside and outside the cylinder. This model is discussed in section \ref{4}, see Fig.\ref{cyl1}.
	
	\indent II. The electrons come from the center, but they cannot cross the outer boundary. The stationary state
	is not established. An electric field is not created inside the cylinder, and the electric current is maintained by a temperature gradient, as in the case of I. The negative charge at the outer boundary increases over time, creating a growing electric field $E_r(t)$ outside the cylinder. This model is discussed in section \ref{5}, see Fig. \ref{cyl2}.
	
	\indent III. Electrons cannot cross any boundary, the total electric charge of the cylinder with a radial heat flow remains zero, but an internal electric field is formed, which negates the electric current. The processes in such a cylinder are discussed in the section \ref{6}, see Fig. \ref{cyl3}.
	
	\indent IV. Electrons freely cross the outer and inner boundaries of the cylinder in such a way that a radial electric field is not created, and it can
	be assumed that $E_r=0$ everywhere. In this case, a continuous supply of electrons is needed near the axis of the cylinder. 
	In our simplified model, this looks artificial, but in a more complex model with a neutron star, something similar could be realized. The idealized one-dimensional
	model is considered in the section \ref{3}, see Fig.\ref{cyl4}.
	
	The heat flow outside the cylinder is completely determined by photon radiation from the surface in Fig. \ref{cyl2} - \ref{cyl3} and partially in Fig. \ref{cyl1}, \ref{cyl4}.
	\bigskip
	\bigskip
	
	\section{Models with open and closed borders}

	\subsection{Model I, where electrons freely cross the outer boundary of the cylinder without electrons coming from the center}\label{4}

\begin{figure}[h!]
		\begin{center}
			\includegraphics[width=0.52\textwidth]{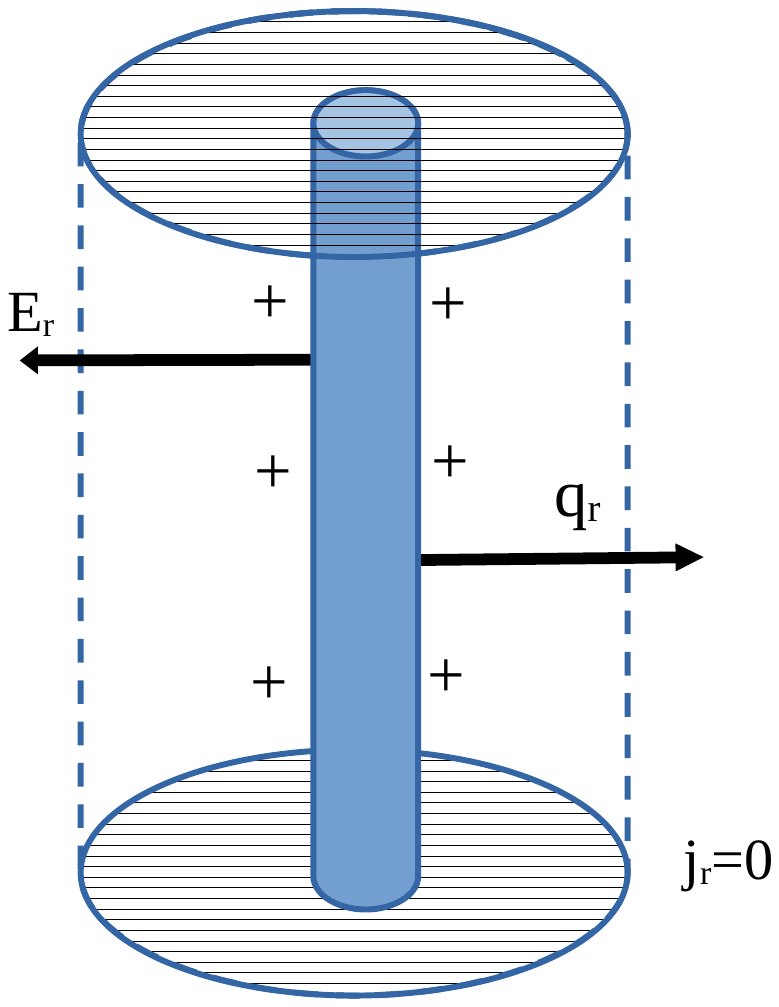}
			\caption{A model of a cylinder with a free outer boundary, without the supply of electrons from the center. The cylinder goes into a stationary state, acquiring a positive electric charge, which creates an internal and external radial electric field and brings to zero the electric current. The heat flow outside the cylinder is determined by the diffusion of photons. The figures here and below are given for the case of a zero magnetic field.}
			\label{cyl1}
		\end{center}
	\end{figure}

	Electrons cross the outer radius of the cylinder under the action of radial heat flow, increasing the positive charge of the cylinder. In this case, the electrons fly out through the outer boundary until the moment when the action of a positive electric charge formed on the inner cylinder does not balance the action of the heat flow, see Fig.\ref{cyl1}.
	The magnitude of such an electric field is determined from the condition $j_r=0$ in
	(\ref{qj1}) with a magnetic field equal to zero. Using the Lorentz gas approximation (\ref{eq2}) to account for the influence of the magnetic field, we obtain at $j_r=0$ the same relationship of the electric field with the temperature gradient. The stationary heat flow in this model is determined from the ratios:
	
	\begin{eqnarray}\label{qj2}
		j_r=j_\varphi=0,   \quad
		E_r= -\frac{\sigma_T}{\sigma_E}\frac{dT}{dr}, \qquad \qquad \nonumber\\
		q_r=
		-\left(\tilde\lambda_T- \lambda_E\frac{\sigma_T}
		{\sigma_E}\right)\frac{dT/dr}{1+(\omega_B\tau_e)^2}
		=-\frac{2}{5}\frac{\tilde\lambda_T dT/dr}
		{1+(\omega_B\tau_e)^2}, \nonumber \\ \qquad
		q_\varphi= q_r\omega_B\tau_e. \qquad \qquad \qquad
	\end{eqnarray}
	Here $Q$ is a constant radial heat flow coming out of the unit length of the cylinder. The radial heat flow per unit area is $q_r = Q/(2\pi r)$.	
	The distribution of temperature and electric field along the radius of the cylinder is written as:
	
	\begin{eqnarray}
		\frac{dT}{dr}=-\frac{5}{2 \tilde\lambda_T}[1+(\omega_B\tau_e)^2] q_r,\qquad \qquad \qquad\\
		E_r= \frac{\sigma_T}{\sigma_E}\frac{5}{2 \tilde\lambda_T}
		[1+(\omega_B\tau_e)^2]q_r=\frac{5\sigma_T}{2 \sigma_E  \tilde\lambda_T}[1+(\omega_B\tau_e)^2] \frac{Q}{2\pi r}.\nonumber
		\label{qj3}
	\end{eqnarray}
	With constant physical parameters along the radius $\sigma_T,\,\lambda_E,\, \tilde\lambda_T,\,\omega_B\tau$  this distribution of the electric field is created by the linear density of the positive charge $\rho_e$ on the axis of the cylinder, with
	\begin{eqnarray}\label{qj4}
		\rho_e = \frac{5\sigma_T}{2 \sigma_E \tilde \lambda_T}[1+(\omega_B\tau_e)^2]\frac{Q}{2\pi},
	\end{eqnarray}
	and zero electric charge density in the cylinder itself.
	We believe that the central cylinder has a very small but finite radius.
	
	\bigskip
	
	\subsection{Model II, where there is a continuous supply of electrons from the central region, but they do not cross the outer boundary of the cylinder\label{5}}
	
	In this case, it is assumed that electrons can penetrate into the cylinder near the central axis, from the central cylinder of a very small radius.
	
\begin{figure}[h!]
		\begin{center}
			\includegraphics[width=0.52\textwidth]{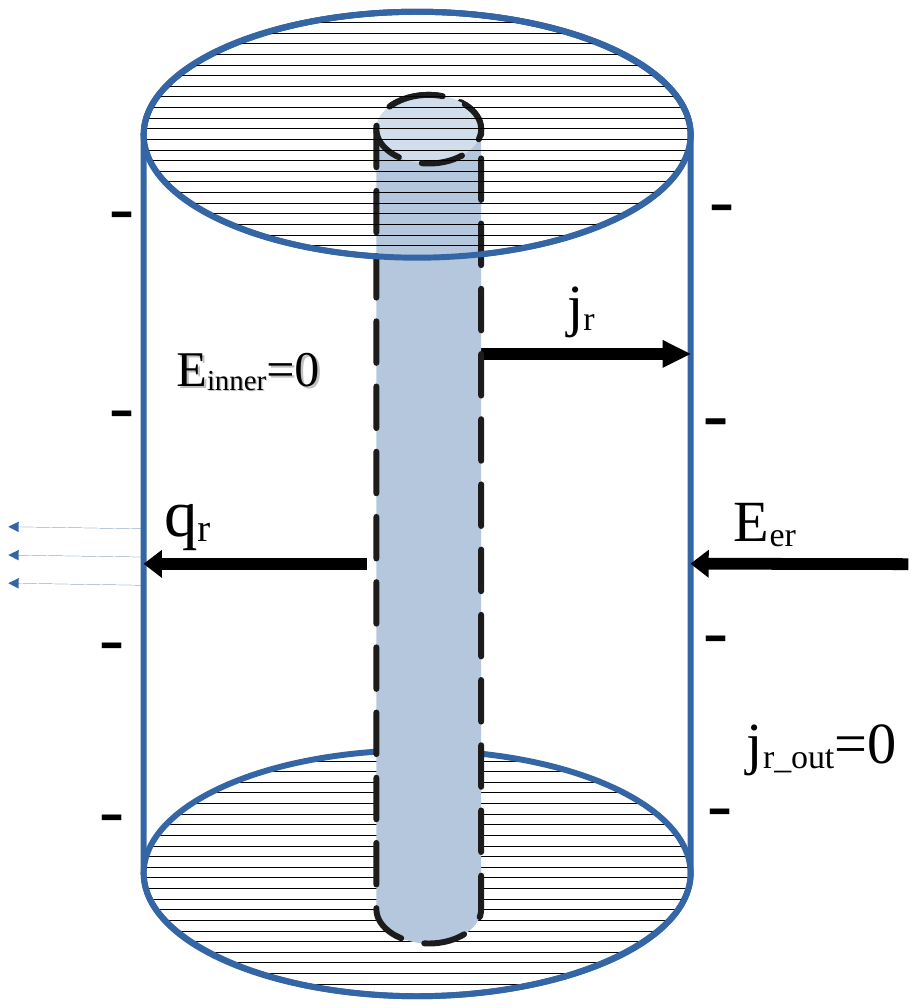}
			\caption{A model of a cylinder with a continuous supply of free electrons from a central source and without crossing an external boundary.  A negative charge is formed at the outer boundary, the external electric field $E_r$ increases over time due to an increase in the surface negative charge, but the internal electric field remains zero.  The heat flow outside the cylinder is determined by the diffusion of photons.}
			\label{cyl2}
		\end{center}
	\end{figure}

\begin{figure}[h!]
		\begin{center}
			\includegraphics[width=0.52\textwidth]{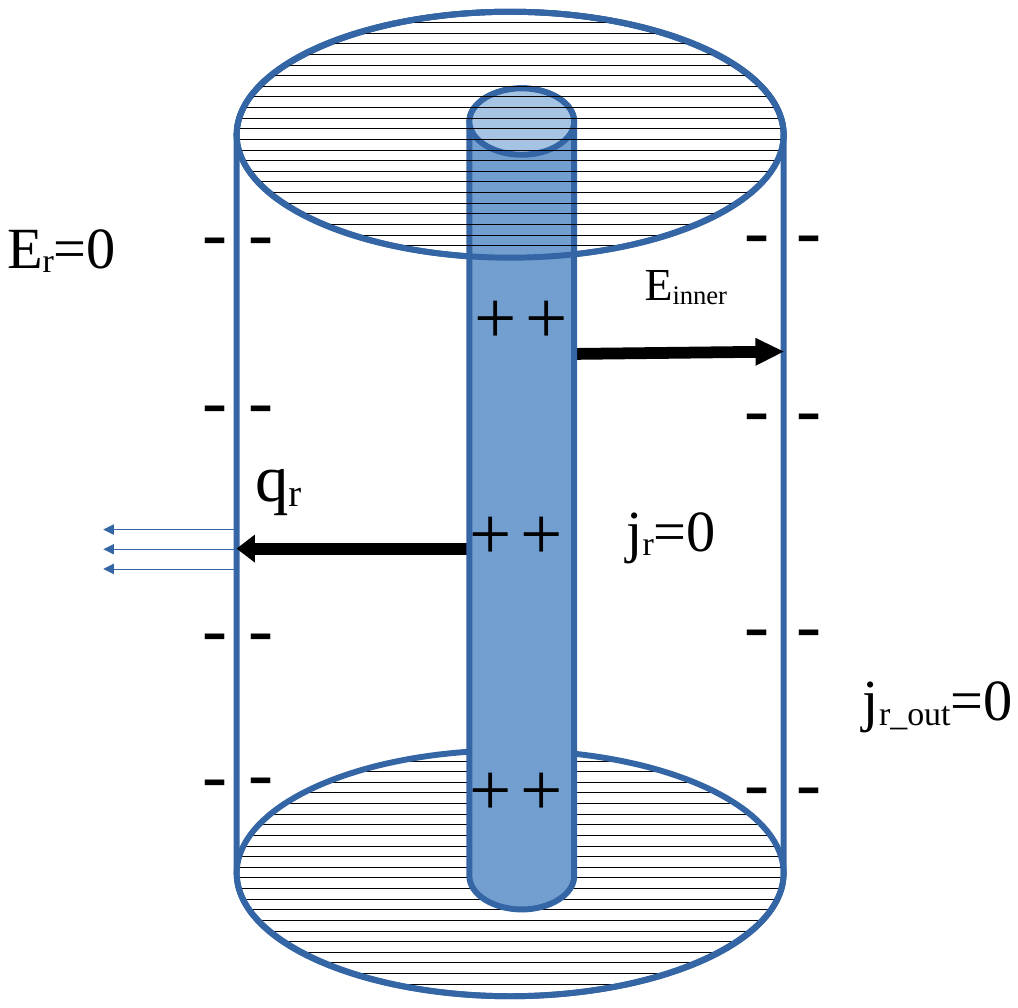}
			\caption{A cylinder model without supply of electrons from a central source, and without penetration through an external boundary. A positive central charge is formed on the axis, and a negative one is formed on the outer boundary. In a stationary state, the external electric field is zero, as is the electric current, due to the zero total electric charge of the cylinder and the balancing of the action of the heat flux and the electric field inside it. The heat flow outside the cylinder is determined by the diffusion of photons.}
			\label{cyl3}
		\end{center}
	\end{figure}

\begin{figure}[h!]
		\begin{center}
			\includegraphics[width=0.52\textwidth]{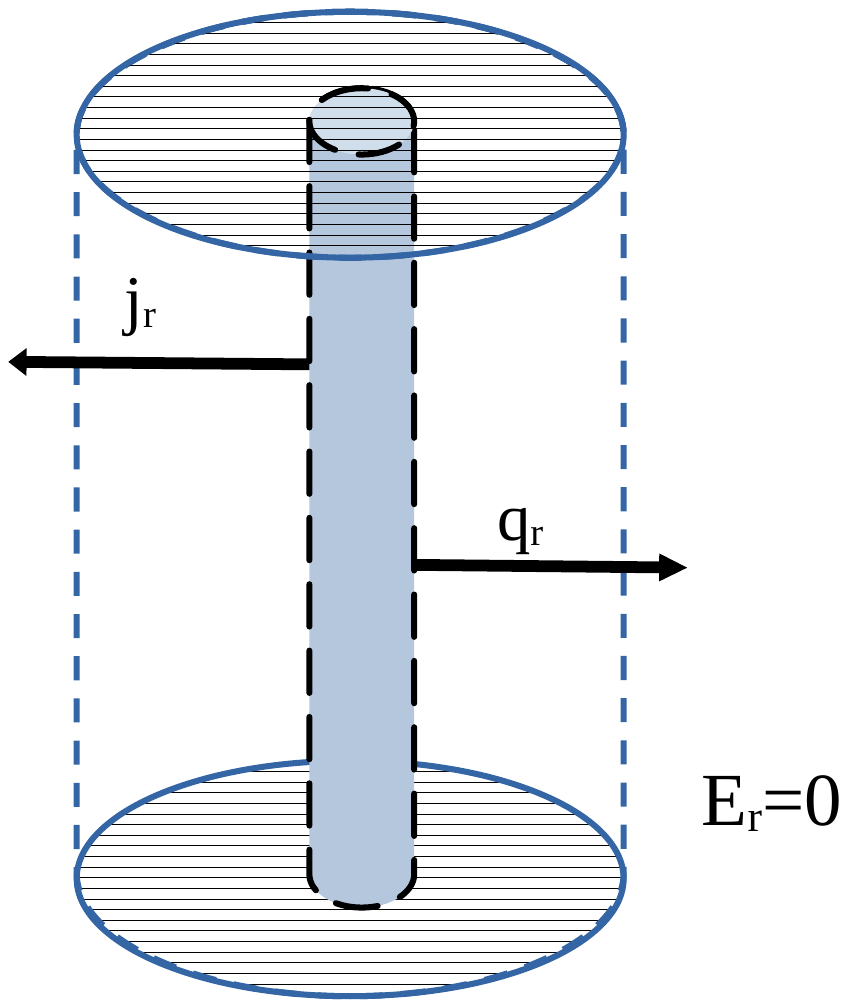}
			\caption{A model of a cylinder with a free outer boundary and a free flow of electrons from a central source from \cite{bkglaph23}. The total radial heat flow inside the cylinder $Q$ is considered a constant value, the electric field $E_r$ is zero. The heat flow outside the cylinder is determined by the diffusion of photons and the flow of electrons.}
			\label{cyl4}
		\end{center}
	\end{figure}

	Stationary state
	in this case, it is not installed. An electric field is not created inside the cylinder, and the electric current is maintained by a temperature gradient, as in the case of I. The negative charge accumulates at the outer boundary, increases over time, creating a growing electric field $E_r(t)$ outside the cylinder, see Fig.\ref{cyl2}. It is obvious that the negative charge of the boundary circle of unit length $\rho_{eb}$ and the strength of the external electric field around the cylinder grow linearly with time, and at zero magnetic field are determined by the relations:
	\begin{equation}
		\rho_{eb}=-\frac{\sigma_T Q}{\tilde\lambda_T}t,\qquad
		E_{er}=-2\frac{\rho_{eb}}{r},
	\end{equation}
	at $r\geq R$.
	
	\bigskip
	
	\subsection{Model III, in which electrons do not cross any boundaries of the cylinder, and its total electric charge remains zero \label{6}}

	In this case, electrons gather near the outer boundary of the cylinder, and the cylinder itself acquires a positive charge, creating an internal electric field that stops the electric current in a stationary state, see Fig. 3. This electric field is determined from the condition $j_i=0$, see
	(\ref{qj1}) for a zero magnetic field. If there is a longitudinal magnetic field, an electric field of the same magnitude arises from the magnetic field, 
	as follows from the simplified consideration of the influence of the magnetic field in
	the Lorentz gas model (ref{eq2}). The heat flow is determined by 
	relations (\ref{qj2}). The difference from case II is 
	the zero value of the total electric charge of the cylinder in
	a stationary state. In this case, it is not charged, unlike
	in case I, when the cylinder acquires a total positive
	electric charge on the inner string, in accordance with 
	(\ref{qj4}). Note that in the absence of an electric current in
	a medium with a temperature gradient, the coefficient of thermal conductivity 
	$\lambda_T$ is 2.5 times less than the coefficient $\tilde\lambda_T$ of 
	\eqref{tlor},\eqref{qj1}, in which
	the condition of absence of electric current is not imposed. Since most stationary objects have no electric current and other diffusion movements, it is customary in the literature to call the coefficient of thermal conductivity specifically the value $\lambda_T=\frac{2}{5}\tilde\lambda_T$ \cite{Chapmen-1952}.
	
	\bigskip
	
	\subsection{Model IV, with free escaping and incoming electrons} \label{3}
	
	In this case, the electrons freely cross the boundary of the cylinder and receive an influx of electrons from the central region. The electric charge density does not develope, and we get $E_i=0$, see Fig. 4. The processes occurring in such a model were analyzed in detail in the article \cite{bkglaph23}, where the role of Hall currents in the formation of the resulting magnetic field in a conducting magnetized cylinder was considered.   Analytical solutions and numerical calculations were obtained for conditions close to the plasma parameters in the crust of neutron stars and for plasma in the laboratory.
	
	\bigskip
	
	\section{Discussion}\label{7}

	A model of an electric generator in the form of a cylinder with a radial heat flow could be designed in a laboratory using a metal or plasma cylinder with a highly heated axis.
	
	A specific feature of this model of an electric generator is the possibility of creating a device in which the electric field increases over time while conserving the parameters of the model.
	The physical limit of the generated field strength is reached when the induced field strength becomes equal to the electron emission threshold. This field becomes stationary and does not change over time. 
	
	A similar situation may occur in a neutron star after its birth. Some layers of a neutron star can become electrically charged, and during accretion, as a result of the influx of external electrons and due to the emission of electrons by an electric field, a neutron star can acquire a non-zero electric charge. 
	The application of this model to obtain realistic neutron star parameters requires
	further consideration.

	Such models are necessary to study the magnetothermal evolution of magnetic and electric fields in neutron stars and white dwarfs, taking into account the anisotropic heat and electric current fluxes caused by magnetic fields and Hall terms in the transfer coefficients.

	\end{document}